# NUMERICAL SIMULATIONS OF GEOMECHANICAL STATE OF ROCK MASS PRIOR TO SEISMIC EVENTS OCCURRENCE – CASE STUDY FROM A POLISH COPPER MINE AIDED BY FEM 3D APPROACH

**Witold Pytel**[1], **Piotr Mertuszka**[1], **Tristan Jones**[2], **Henryk Paprocki**[3]

[1]KGHM CUPRUM Ltd. Research and Development Centre, **Poland**
[2]Luossavaara-Kiirunavaara AB, **Sweden**
[3]KGHM Polska Miedź S.A., Lubin Mine, **Poland**

**ABSTRACT:**

*The biggest risk to operations in Polish copper is high-energy seismic tremors where hypocenters are located within the main roof strata. These phenomena occur often in the areas which are virtually free of tectonic disturbances. They are extremely difficult to both predict and analyze due to their short duration and unpredictable time of occurrence. The objective of the presented analysis was to determine the overall stress/deformation states within the considered mining panel prior to the selected archival seismic event and afterwards to judge how the progress of mining could affect the occurrence of instability in the roof strata rock mass. Geomechanical problem solution and results visualization were based on the NEi/NASTRAN computer program code utilizing FEM in three dimensions.*

## 1. INTRODUCTION

The Lower Silesian Copper District is located in south-west part of Poland and this is one of the most important mining areas in Europe, extracting over 30 million tons of copper ore per year. Copper ore is excavated in three underground mines belonging to KGHM: the Lubin, Polkowice-Sieroszowice and Rudna mines. The stratoidal deposit is characterised by variable thickness (from 0.4 to 26 m), small inclination (approximately 4°), and varying lithological profile. The room-and-pillar mining system is a dominant technology utilized in underground copper mines in Poland. This mining method is well suited to the hardness of the local rock and is proven to be highly adaptable and suitable to the local mining and geological conditions [1].

With the increasing depth of exploitation and higher variability of rock-mass characteristics within roof strata, it was observed that increasing stress created more difficulties in the mining process [2]. The biggest risk to the mining operations is created by high-energy tremors with hypocenters located within the main roof strata, about 40-200 m above the excavated copper ore body. They are extremely difficult to both predict and to analyze due to their short duration and unpredictable time of occurrence [3]. The strongest tremors, reaching seismic energy levels over $10^9$ J, can be considered as small earthquakes and are quite often associated with rockbursts. Due to the dynamic nature of the rock mass failure, some of these rockbursts can cover the working areas, becoming a hazard to both crew and equipment safety.

The aim of the paper was to determine the overall stress/deformation states within the considered mining panel immediately prior to selected historic seismic events and subsequently to judge how the progress of mining could affect the occurrence of instability in the roof strata. The geomechanical problem solution and results visualization were based on the NEi/NASTRAN computer program code utilizing 3D finite elements method.

## 2. SEISMICITY WITHIN THE CONSIDERED MINING PANEL

The XIII/4 mining panel is located in the central part of the Lubin – Małomice mining area, about 3 km to the east of the L-I and L-II shafts on the depth between 728 m and 785 m beneath the surface. The form of the deposit is classified as stratoidal and single-level. The ore bed includes grey Rotliegend sandstones and copper-bearing lower-Zechstein shales. The deposit is extended in the NW-SE direction and declines about 3° in the NE direction.

The specific geologic structure of the overburden, which is typical in KGHM's copper mines, favours the occurrence of strong seismic events. The material of the strata is mostly characterised by high strength and low deformability. From the geomechanical point of view, the occurrence of dynamic events is caused by high stress-strain levels within the material surrounding mine workings, leading to a low value of the bearing capacity margin. This means that even a slight additional load in the form of adverse stress changes – according to the accepted strength hypothesis – may lead to instability in a certain area of the workings.

Experiences gained from previous observations lead to the conclusion that the seismicity level depends on progress of mining in a particular panel, and on the geologic and mining conditions. This is why seismicity varies over the years. Figure 1 shows the seismic activity in the all KGHM mines starting from 1990. Tremors with the seismic energy lower than $10^3$ J were omitted.

Based on Figure 1 one may conclude that the number of tremors have generally increased in the past years. General trend of number of events could be approximated by linear function with the relatively high coefficient of determination $R^2=0,8586$. The total annual energy emission from mining-induced tremors increased extremely between 1990 and 2000 when seismic energy reached the peak value. High level of seismic activity remained stable over the next 2 years and then decreased significantly. This situation stabilised in 2007. The seismic activity level has remained relatively low and stable from 2007 until 2016.

The subject of the case of study, the XIII/4 mining panel, is one of the panels in the Lubin mine. The deposit within this mine is characterized by the frequent occurrence of tectonic discontinuities within the orebody, which can create difficulty during mining operations, even leading to abandonment in some case

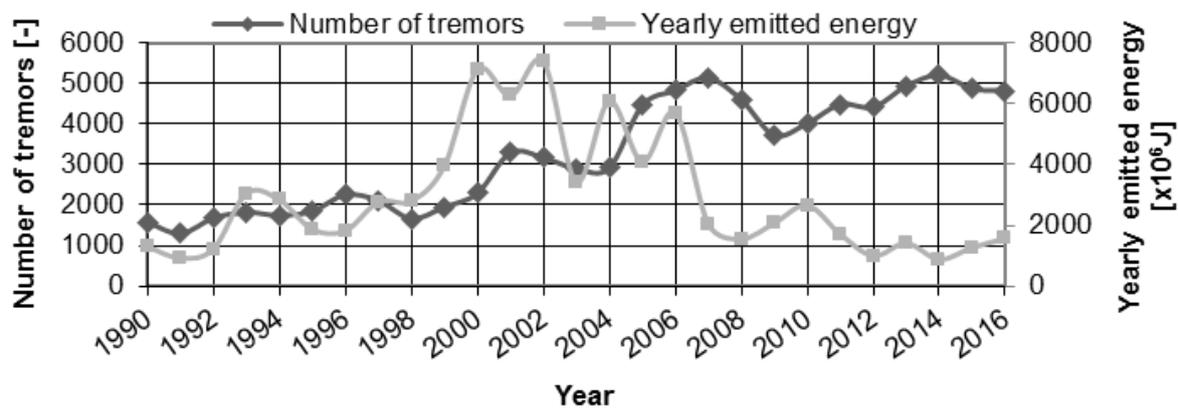

Figure 1. Number and energy of tremors recorded between 1990 and 2016 in KGHM mines

The inventory of the seismic events from the Lubin mine between 2007 and 2016 (see Fig. 2) indicates that significant fluctuation can be observed in practically all energy classes, especially in the case of events with the seismic energy greater than $10^6$ J. When considering exclusively tremors with the energy greater than $10^7$ J, sudden increase of the number of tremors of this energy class may be observed.

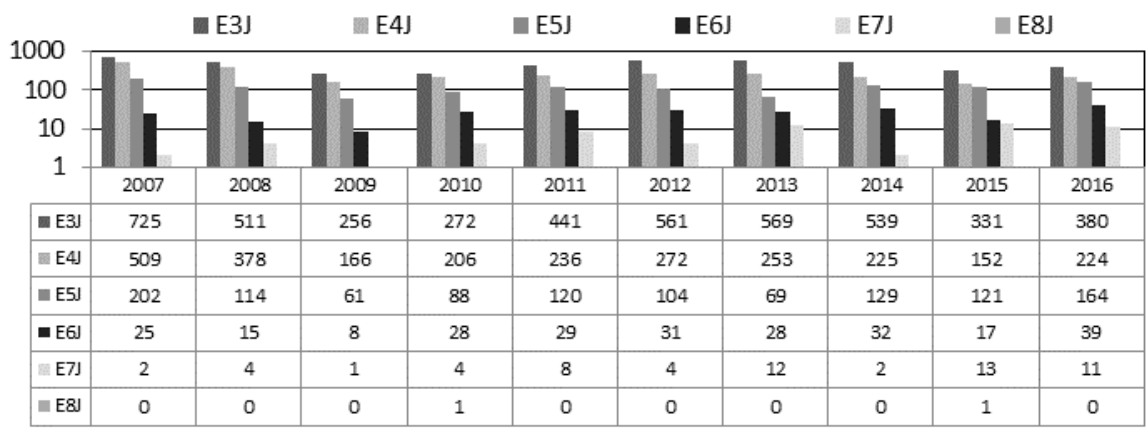

|     | 2007 | 2008 | 2009 | 2010 | 2011 | 2012 | 2013 | 2014 | 2015 | 2016 |
|-----|------|------|------|------|------|------|------|------|------|------|
| E3J | 725  | 511  | 256  | 272  | 441  | 561  | 569  | 539  | 331  | 380  |
| E4J | 509  | 378  | 166  | 206  | 236  | 272  | 253  | 225  | 152  | 224  |
| E5J | 202  | 114  | 61   | 88   | 120  | 104  | 69   | 129  | 121  | 164  |
| E6J | 25   | 15   | 8    | 28   | 29   | 31   | 28   | 32   | 17   | 39   |
| E7J | 2    | 4    | 1    | 4    | 8    | 4    | 12   | 2    | 13   | 11   |
| E8J | 0    | 0    | 0    | 1    | 0    | 0    | 0    | 0    | 1    | 0    |

Figure 2. Number of seismic events of the particular energy classes recorded within the Lubin mine between 2007 and 2016

However, when considering the relationship between the event count and the total energy released in each energy class, one may conclude that most of energy comes from large events with energy magnitudes greater than $10^6$ J. It should be also noted, that the total energy of tremors is inversely proportional to frequency of events occurrence. The differences between the energy released in any two subsequent years varied significantly, from 12% up to 343%. This is the best example of how dynamic events' occurrence is unpredictable.

The seismic activity in this panel has remained at a very high level from the beginning of mining works in this area. The frequency of tremor occurrence and quarterly released energy increased almost linearly since 2014. The XIII/4 mining panel was active in the light of seismic activity, both spontaneous and blasting-induced, which allows the determination of the relationship between the stress/deformation states within the analyzed mining panel both prior to and after a specific seismic event. Based on this relationship it is possible to judge how the progress of mining could affect the likelihood of the occurrence of instability in the rock mass.

From a geometric point of view, the locations of almost all recorded high-energy tremors were determined to be in a single, straight line (Fig. 3). This indicates, that the mining-geological conditions encountered on the right side of the contact between undisturbed rock mass and the excavated workings was the main source of the geomechanical type of instabilities.

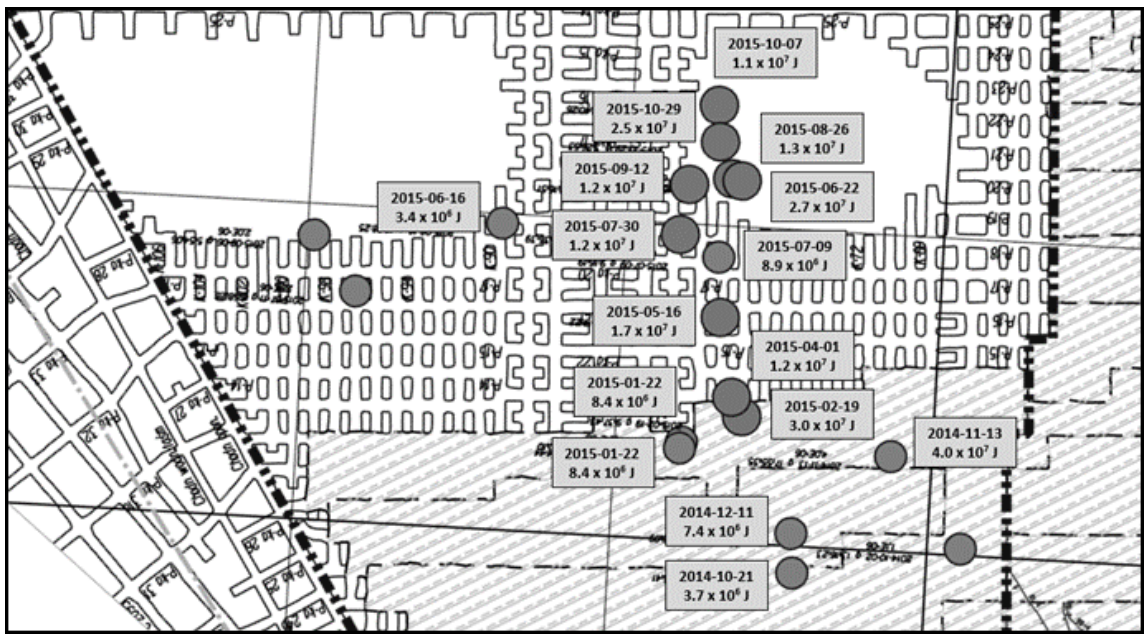

Figure 3. Locations of high-energy tremors within XIII/4 mining panel between 2014 and 2015

### 3. GEOMECHANICAL ANALYSIS BASED ON FEM MODELLING

Twenty tremors with an energy level of $10^7$ J occurred between February 2015 and November 2016 within the considered mining panel, 10 of which caused serious damage to the mine workings. For the purpose of presented paper, three cases of tremors of which full documentation was available, were examined. Computer simulations were performed using a numerical model based on the geometry of the mine workings at the time when the following seismic events occurred (Fig. 4):

- June 22, 2015 – seismic energy of $E_s$ = 2.7 x $10^7$ J with excavations' geometry described later as "Event 1".
- May 11, 2016 – seismic energy of $E_s$ = 1.9 x $10^7$ J with excavations' geometry described later as "Event 2".
- November 3, 2016 – seismic energy of $E_s$ = 7.6 x $10^6$ J with excavations' geometry described later as "Event 3".

Geomechanical problem solution and results visualization were based on the NEi/NASTRAN computer program code utilizing FEM in three dimensions [4]. It was assumed that all of the materials reveal linear elastic characteristics, except for rocks comprised within pillars which exhibit elastoplastic behaviour with strain softening [5]. The boundary conditions of the entire numerical models were described by displacement-based relationships. A non-linear calculation procedure was applied, including an adaptive phase (elastic solutions with successive modification of overloaded pillars and an iterative procedure for selection of pillars for which the vertical load $\sigma_z$ was greater than the critical value $\sigma_p$, and subsequently using a constant residual load $\sigma_r$ instead of such pillars) and a final phase (verification of safety factors within the roof layers). The displacement boundary conditions were formulated as follows: no vertical displacements of the bottom layer in the model and no movements in the direction perpendicular to the lateral walls. The external load was divided into the following two groups: the self-weight load of the tertiary and quaternary formations and the self-weight load of the other rocks represented by material characteristics, including the density (gravitation).

As a basic physical model for the problem, the multi-plate overburden model has been accepted. It was assumed that the overburden strata consists of several homogeneous rock plates reflecting the real lithology in the area and that the technological and remnant pillars work effectively within a post-critical phase (elastic-plastic with strain softening behaviour) [6], [7]. To illustrate the effect of mining face geometry, the XIII/4 mining district has been modelled in 3D FEM code. A mining system with roof deflection was assumed. The averaged geological data over the considered area and the estimated rock mass parameters are given in Table 1.

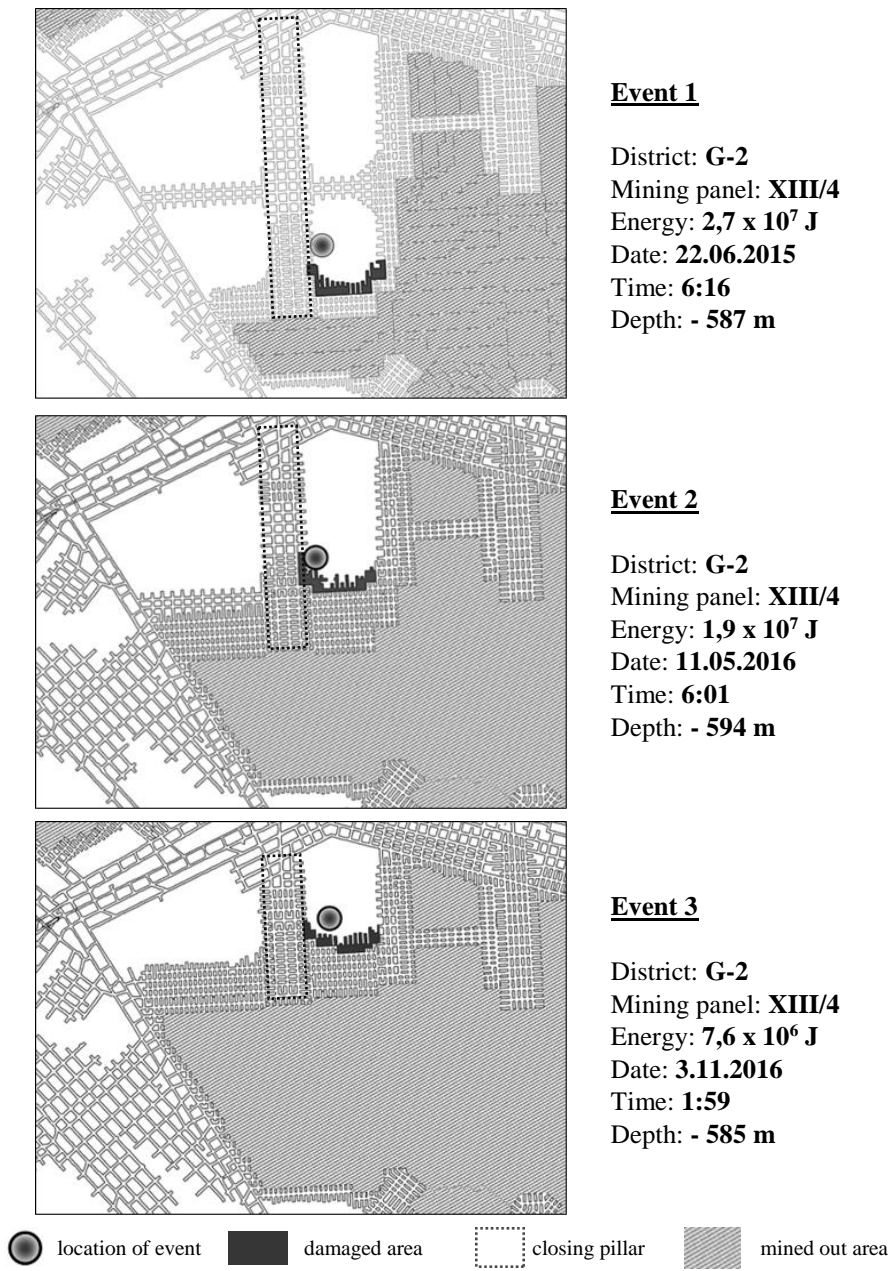

**Event 1**

District: **G-2**
Mining panel: **XIII/4**
Energy: **2,7 x 10^7 J**
Date: **22.06.2015**
Time: **6:16**
Depth: **- 587 m**

**Event 2**

District: **G-2**
Mining panel: **XIII/4**
Energy: **1,9 x 10^7 J**
Date: **11.05.2016**
Time: **6:01**
Depth: **- 594 m**

**Event 3**

District: **G-2**
Mining panel: **XIII/4**
Energy: **7,6 x 10^6 J**
Date: **3.11.2016**
Time: **1:59**
Depth: **- 585 m**

○ location of event    ■ damaged area    ▢ closing pillar    ▨ mined out area

**Figure 4. Location of the considered seismic events with the geometry of mining drifts**

Table 1. Geological data in the vicinity of considered area

| Rock type | Thickness (m) | $\sigma_c / \sigma_{cm}$ (MPa) | $\sigma_t / \sigma_{tm}$ (MPa) | $E_s / E^{(r)}$ (MPa) | Poisson's ratio ($\nu$) | Level |
|---|---|---|---|---|---|---|
| Clayey Shale | 12.2 | 22.5 / 1.56 | 1.7 / 0.006 | 13500 / 3375 | 0.18 | ROOF STRATA |
| Anhydrite | 136.8 | 88.5 / 17.1 | 6.25 / 0.21 | 55500 / 13875 | 0.18 | |
| Limestone | 13.8 | 174.0 / 92.1 | 9.9 / 1.3 | 52500 / 13125 | 0.24 | |
| Dolomite II | 10.8 | 119.3 / 37.0 | 8.2 / 0.49 | 57100 / 14275 | 0.24 | |
| Dolomite I | 9.2 | 146.7 / 61.4 | 9.9 / 0.85 | 51400 / 12850 | 0.24 | |
| Copper ore | 2.8 | 40 | 2.7 | 13000 / 6500 | 0.17 | EXTRACTION RANGE |
| White Sandstone | 10 | 34.4 | 1.9 | 11600 / 5800 | 0.17 | FLOOR STRATA |
| Quartz Sandstone | 290 | 17.9 | 0.9 | 5100 / 2550 | 0.13 | |

- $\sigma_{cm}$, $\sigma_{tm}$ compression and tension strengths for rock mass, assessed according to Hoek's [8] approach

Numerical experiments permitted the determination of the overall stress/deformation states which were later used for quantitative characterization of the actual level of safety, using the indicator called safety factor, related to different failure criterions. In this paper safety factor $F_{HB}$ spatial distribution has been calculated utilizing the Hoek-Brown theory (see [6]):

$$\sigma_1 = \sigma_3 + \sqrt{A\sigma_3 + B^2} \quad (1)$$

$$\text{or } \tau_m = \frac{1}{8}\left[-A \mp \sqrt{A^2 + 16(A\sigma_m + B^2)}\right] \quad (2)$$

Safety factor may be then presented as follows:

$$F_{HB} = \frac{\frac{1}{8}\left[-A \mp \sqrt{A^2 + 16(A\sigma_m + B^2)}\right]}{\tau_m} \quad (3)$$

where:

$A = \frac{R_c^2 - R_r^2}{R_r}, B = R_c$;

$\sigma_p$ – average value of principal stresses;

$\tau_m$ – maximum shear stress in specific location of the rock mass.

The values of safety factor smaller than 1 indicate the likely occurrence of instability in the rock mass. The selected results of calculations are shown in Fig. 5-7 as contours of safety factor at 5.2 m (dolomite I), 9.7 m (dolomite I), 14.7 m (dolomite II), 20.1 m (dolomite II), 26.25 m (limestones), 33.15 m (limestones) above the roof strata for considered events.

From the presented geomechanical analysis one may conclude that safety factors will reach the lowest value at levels closer to the excavation, especially in mined-out areas (roof fall hazard). Furthermore, it was found that in the all three considered cases, safety factor values $F_{HB}$ << 1 are located both in mined-out zones as well as in working areas. The lowest values of safety factors were calculated in the vicinity of so called *closing pillar* in the immediate roof strata within Dolomite I and Dolomite II, i.e. 2.8-22.8 m above the floor level of the excavations. In addition, all recorded high-energy seismic events were developed within that height range.

With respect to geometry, in general, a series of high-energy tremors occurred on the right side of the mining front. As the mining front progressed, they were observed at relatively regular time-space intervals, almost in a one straight line. This suggests, that they were caused by the geometry of the rock mass shaped in the form of a sharp corner directed to yielded pillars.

The results of calculations coincide with the actual conditions, i.e. the indicated areas of increased risk of instability coincide with the actual range of damages around the workings. The determined area of possible instability occurrence contained the location of all considered high-energy tremors.

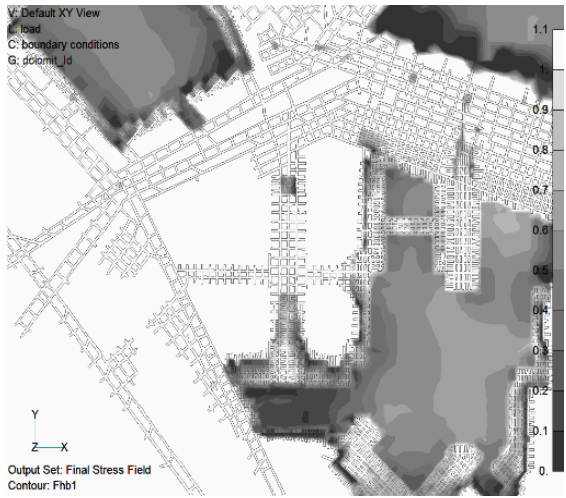
z = 5.2 m (dolomite I)

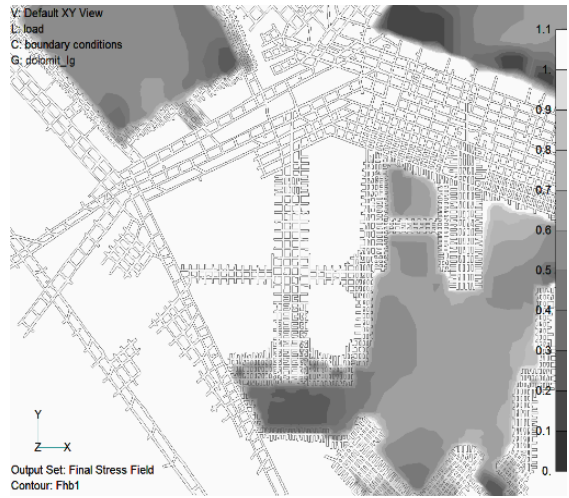
z = 9.7 m (dolomite I)

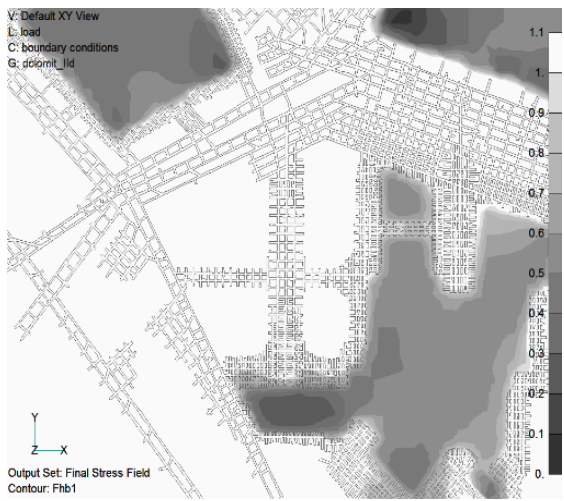
z = 14.7 m (dolomite II)

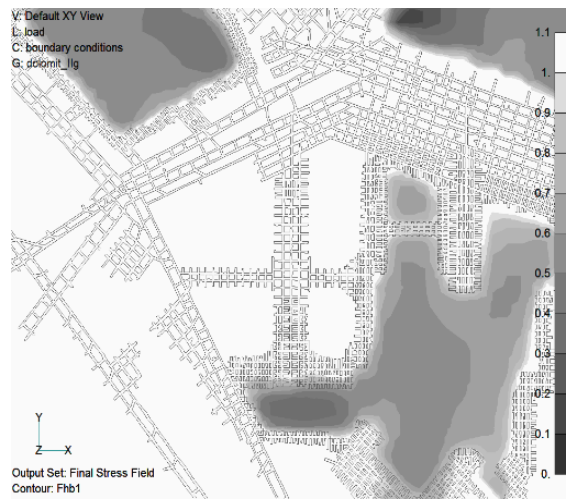
z = 20.1 m (dolomite II)

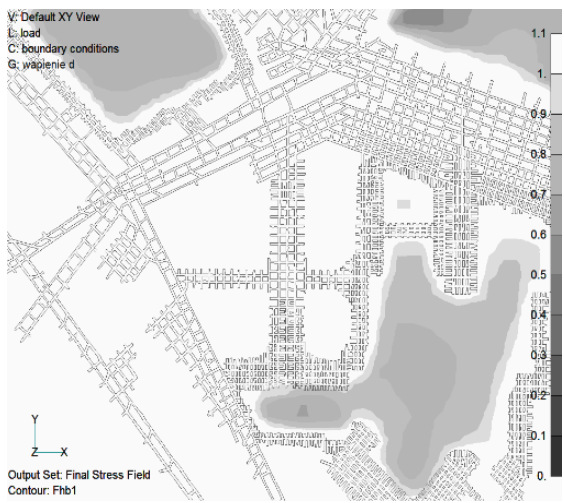
z = 26.25 m (limestones)

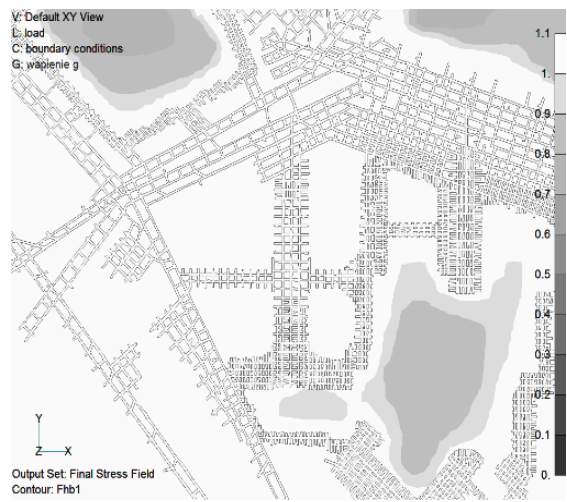
z = 33.15 m (limestones)

**Figure 5. Contours of calculated values of safety factor in the roof strata (event 1)**

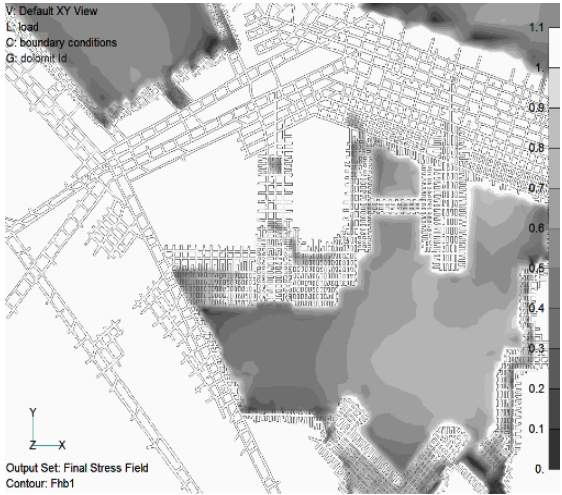
z = 5.2 m (dolomite I)

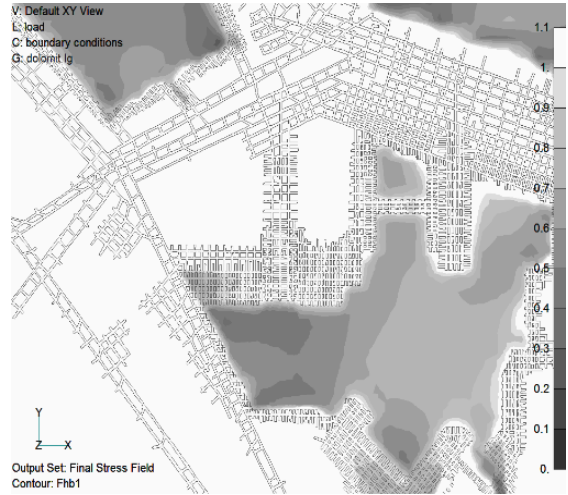
z = 9.7 m (dolomite I)

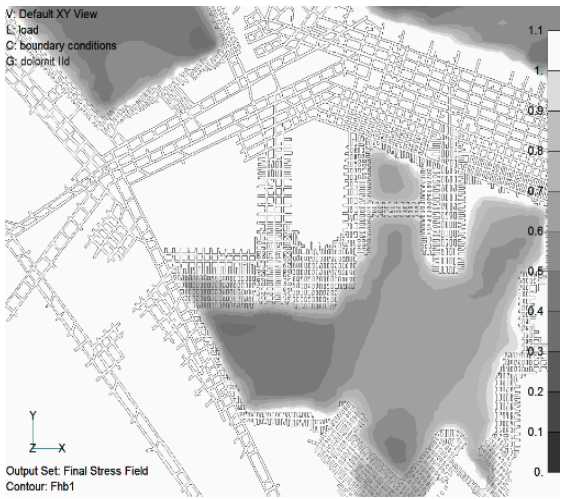
z = 14.7 m (dolomite II)

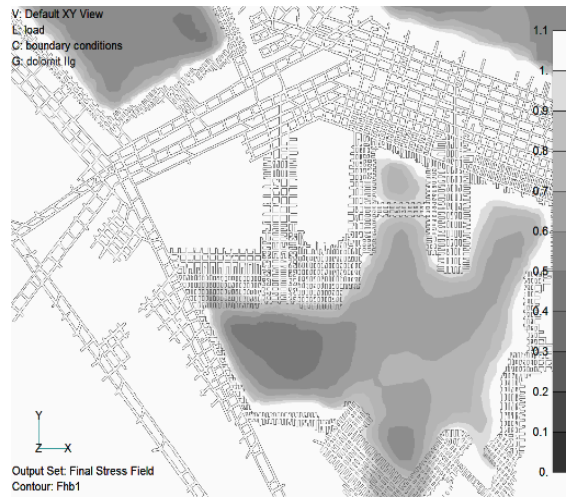
z = 20.1 m (dolomite II)

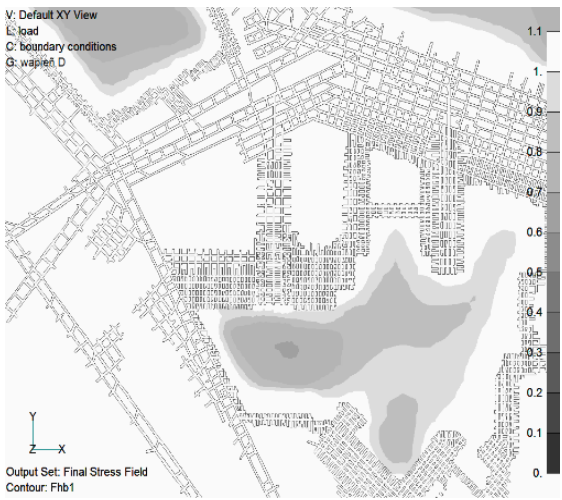
z = 26.25 m (limestones)

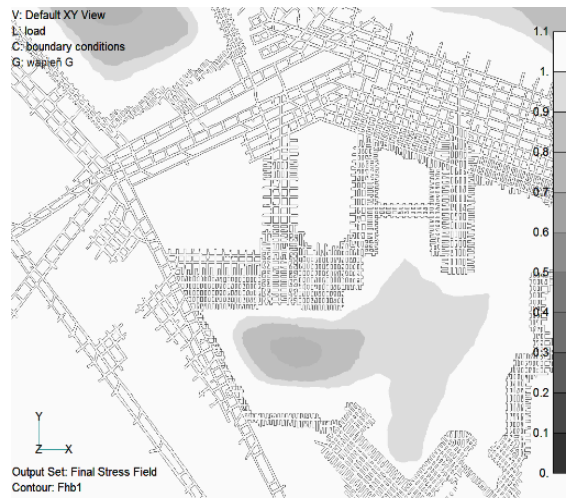
z = 33.15 m (limestones)

**Figure 6. Contours of calculated values of safety factor in the roof strata (event 2)**

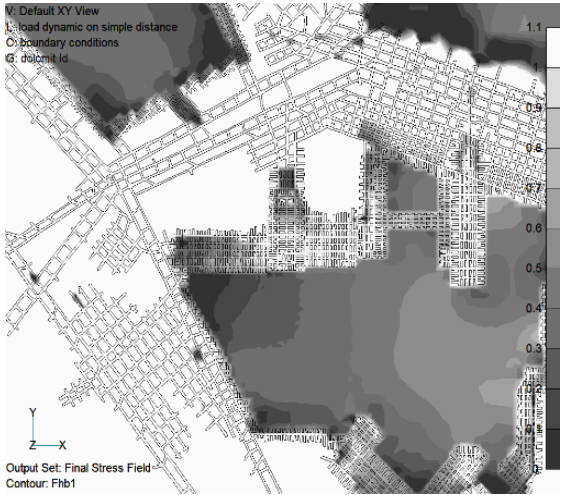
z = 5.2 m (dolomite I)

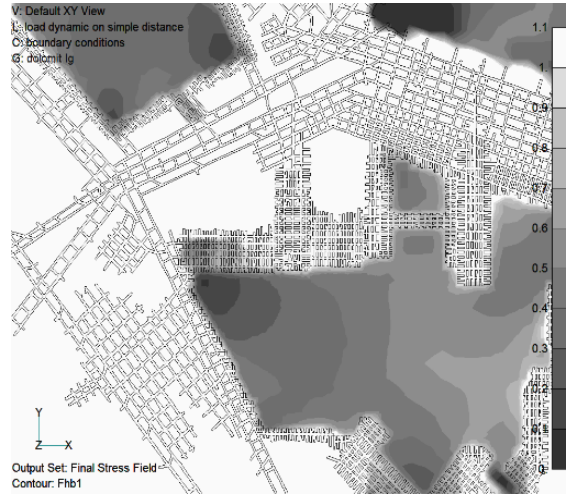
z = 9.7 m (dolomite I)

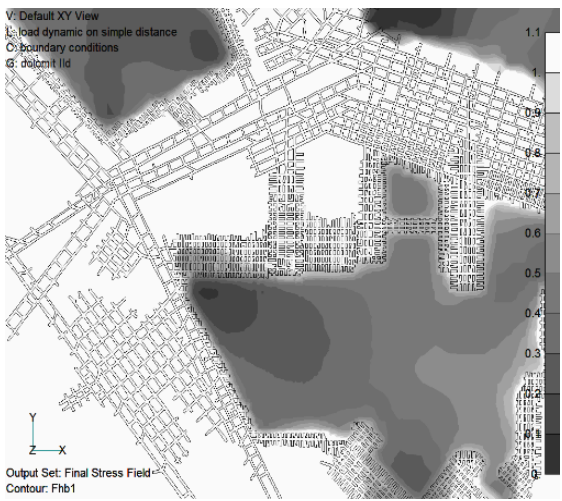
z = 14.7 m (dolomite II)

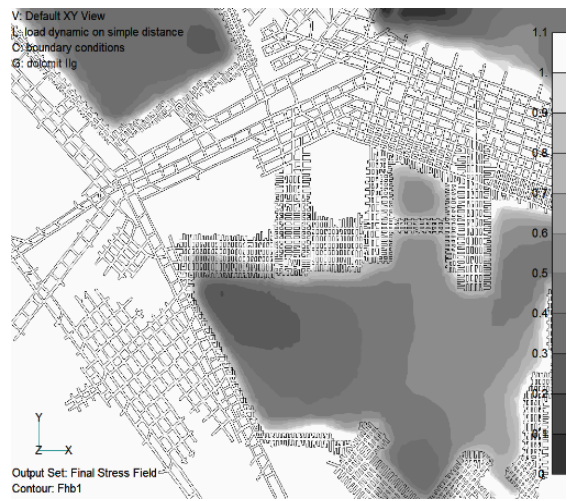
z = 20.1 m (dolomite II)

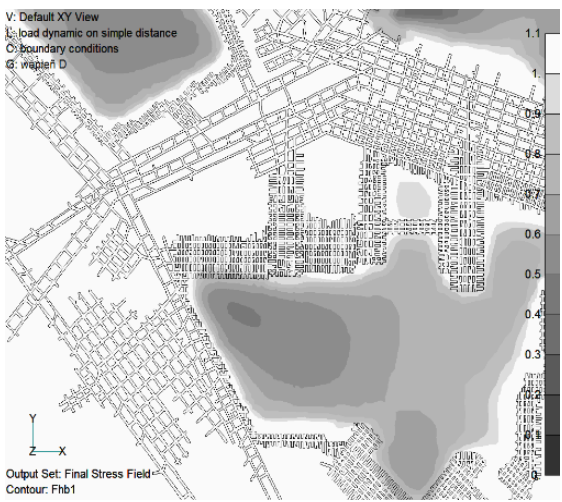
z = 26.25 m (limestones)

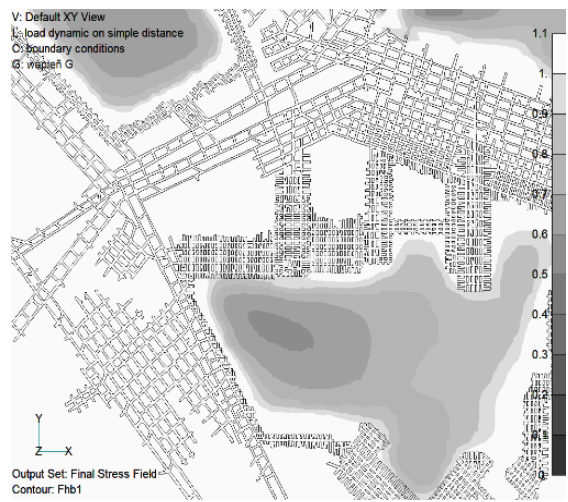
z = 33.15 m (limestones)

**Figure 7. Contours of calculated values of safety factor in the roof strata (event 3)**

## 4. CONCLUSIONS

The presented numerical models, based on the classic finite element method, is a very efficient 3D tool to identify areas of the main roof strata in underground mines that are weak/prone to potential seismicity and rock bursting. The numerical models were validated based on locations of known stress relief events. The knowledge about the areas in which the safety margins indicate that an instability event occurrence is very likely makes it possible to verify the results obtained using the developed numerical model. The applied numerical methods have revealed themselves to be very effective in the verification of the "primary" conditions, i.e. stress and deformations within the analyzed area of the rock mass.

According to the above presented analysis one may conclude that the precise identification of overstressed areas is possible using these numerical tools. This method may be applied as a routine procedure while assessing rock mass behaviour, and possibly in the future as a rockburst prevention method as well.

## ACKNOWLEDGEMENTS

**This paper has been prepared through the Horizon 2020 EU project on "Sustainable Intelligent Mining Systems (SIMS)", Grant Agreement No. 730302.**